\documentclass[english]{article}

\usepackage{textcomp,gensymb}
\usepackage{amsrefs}

\usepackage{spconf,graphicx}
\usepackage{amsmath}
\usepackage{standalone}

\usepackage{graphicx}
\usepackage{subcaption}
\usepackage{cite}
\usepackage{listings}
\usepackage{booktabs,multirow}

\usepackage{import}

\usepackage{pgf}
\usepackage{enumitem}   
\usepackage{graphicx}
\usepackage{float}

\newcommand{\RNum}[1]{\uppercase\expandafter{\romannumeral #1\relax}}

\newcommand{\argmin}{\arg\!\min}

\title{estimation of optimal encoding ladders for tiled 360$\degree$ VR video in \\adaptive streaming systems}
\name{Cagri Ozcinar, Ana De Abreu, Sebastian Knorr, and Aljosa Smolic
\address{Trinity College Dublin (TCD), Dublin 2, Ireland.}}
\begin{document}

%   \begin{table*}[]
%     \begin{minipage}{.4\textwidth}
%       \centering
%       \begin{tabular}{cccc}
%       \toprule 
%       & \multicolumn{3}{c}{QP}\\
%     \cmidrule(l){2-4}
%         &$\Phi$ &$\Psi$ &$\Omega$\\
%         \midrule 0,9509 & 1,3645\\
%         2,3153 & 1,3644\\
%         \bottomrule
%       \end{tabular}
%     \end{minipage}
%     \begin{minipage}{.4\textwidth}
%       \centering
%       \begin{tabular}{cc}
%         \toprule Tempo (s) & Velocidade (m/s)\\
%         \midrule 0,2733 & 5,3763\\
%         0,9496 & 5,3763\\
%         \bottomrule
%       \end{tabular}
%     \end{minipage}
%   \end{table*}

% make the title area
\maketitle

\begin{abstract}
Given the significant industrial growth of demand for virtual reality (VR), 360$^{\circ}$ video streaming is one of the most important VR applications that require cost-optimal solutions to achieve widespread proliferation of VR technology. Because of its inherent variability of data-intensive content types and its tiled-based encoding and streaming, 360$^{\circ}$ video requires new encoding ladders in adaptive streaming systems to achieve cost-optimal and immersive streaming experiences. In this context, this paper targets both the provider's and client's perspectives and introduces a new content-aware encoding ladder estimation method for tiled 360$^{\circ}$ VR video in adaptive streaming systems. The proposed method first categories a given 360$^{\circ}$ video using its features of encoding complexity and estimates the visual distortion and resource cost of each bitrate level based on the proposed distortion and resource cost models. An optimal encoding ladder is then formed using the proposed integer linear programming (ILP) algorithm by considering practical constraints. Experimental results of the proposed method are compared with the recommended encoding ladders of professional streaming service providers. Evaluations show that the proposed encoding ladders deliver better results compared to the recommended encoding ladders in terms of objective quality for 360$^{\circ}$ video, providing optimal encoding ladders using a set of service provider's constraint parameters.
\end{abstract}


% \vspace{-0.8em}
\section{Introduction}
% % \vspace{-0.8em}
\label{intro}

Recent years have witnessed a significant industrial investment in virtual reality (VR) technology that has motivated technical developments of graphic cards and head-mounted displays (HMDs)~\cite{vrmarket}. Currently, the video technology field is evolving toward providing immersive VR experiences using 360$^{\circ}$ video streaming. 360$^{\circ}$ video is captured with omnidirectional camera arrays and the individual camera views are projected onto a sphere. For backward-compatibility purposes with the existing video coding standards and streaming pipelines, the spherical videos are mapped onto a planar surface using projection techniques, such as equi-rectangular projection (ERP). ERP videos contain full panoramic 360$^{\circ}$ horizontal and 180$^{\circ}$ vertical views of the scene.
% , projecting back into a 3D sphere at the time of rendering.

360$^{\circ}$ video streaming is significantly challenging owing to its resource-intensive encoding and storage requirements to cope with the very high resolution of its representation. As the VR end-user can only view the field of view (FoV) of the display device (\textit{e.g.,} HMD, smartphone, tablet or laptop), called viewport, very high resolution of 360$^{\circ}$ video (\textit{e.g.,} 8K$\times$4K ERP) is required for transmission in order to achieve high-quality and seamless video streaming experiences. To reduce both the bitrate consumption of the end-user and the visual distortion of the viewport, 360$^{\circ}$ video frames can be divided into self-decodable regions \cite{heymann2005representation,Grunheit2002du}, namely, tiles.

\begin{figure}
\resizebox{\linewidth}{!}{
        \centering
         \includegraphics[width=1\linewidth]{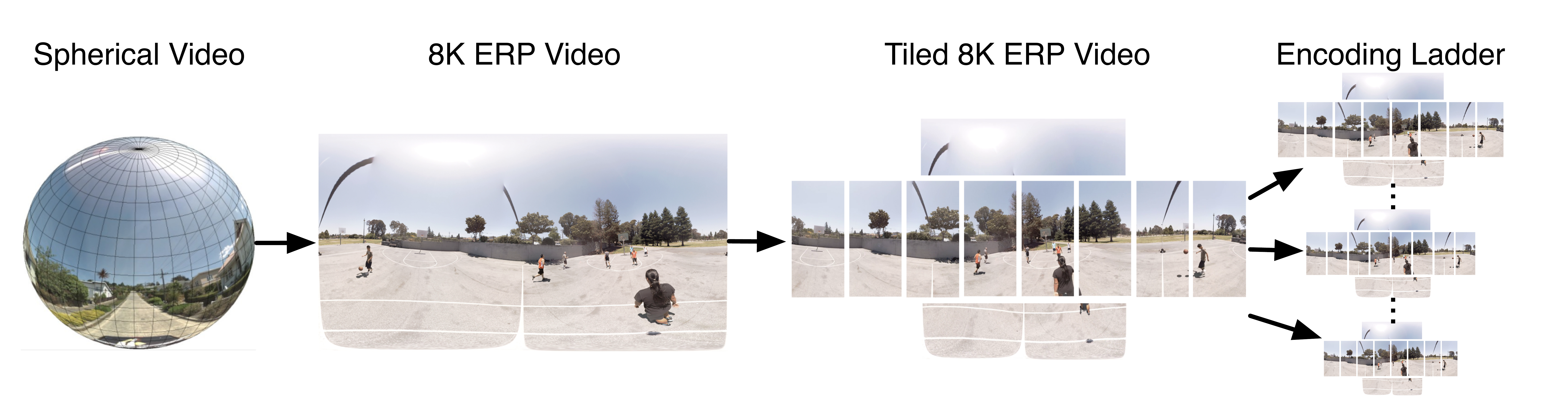}}
         \vspace{-1.5em}
        \caption{Overview of the different formats and representations.}
         \vspace{-1.5em}
        \label{360-projections}
\end{figure}

To deliver the tiled 360$^{\circ}$ videos to the end-user devices, adaptive streaming systems, such as MPEG-dynamic adaptive streaming over HTTP (DASH)~\cite{Isoiec230091undatedlt}, provide smooth 360$^{\circ}$ video streaming experiences, but still require high encoding and storage costs for the tiled 360$^{\circ}$ video. The \emph{spatial relationship description} (SRD)~\cite{Niamut2016hd} can be used with DASH systems where the 360$^{\circ}$ video stream is divided into tiles. In the SRD, each 360$^{\circ}$ video is divided into a set of tiles that includes different \emph{bitrate levels} of the tiled video. Different bitrate levels share the same video content but are encoded using various settings, such as the resolution and the target bitrate for encoding. Each different version is called a \emph{representation}, and a set of representations for the video content forms the \emph{encoding ladder} which is requested by the DASH client to play the tiled 360$^{\circ}$ video. However, encoding and accumulating a large combination of representations for each video content might cover a broad range of network bandwidths such that the end-users can request video streams of appropriate bitrates, and thus it requires high encoding and storage costs~\cite{janEncoding}. Fig.~\ref{360-projections} illustrates the different stages from the spherical projection to the encoding ladder with the different representations of the ERP video.

%\begin{figure*}
%        \centering
        % \includegraphics[trim={0.2cm 0.2cm 0cm 0.2cm},clip,width=\linewidth]{figures/Add_fig.png}
%         \includestandalone[mode=buildnew,width=0.8\linewidth]{figures/Figure2_v1}
        % \vspace{-0.5em}
%        \caption{Overview of the different formats and representations in the processing chain.}
        % \vspace{-1.5em}
%        \label{360-projections}
%\end{figure*}%
% \vspace{-0.3em}

To tackle this problem, cost-optimal encoding ladders are needed for service providers to deliver tiled 360$^{\circ}$ video content and satisfy network bandwidths. In fact, tiled 360$\degree$ video provides different rate-distortion (RD) performance compared to the traditional video content due to different characteristics of both. In particular, tiling affects the coding efficiency, because redundancy cannot be exploited over tiles. Furthermore, given its 2D projection for encoding (\textit{e.g.,} ERP), each tile of the 360$\degree$ video has a different level of contribution for the overall 360$\degree$ video viewing quality due to stretching effects caused by the projection~\cite{qomexAna,Jvet2016tw}. To this end, new encoding ladder configurations are required for the tiled 360$\degree$ videos to provide cost-optimal video streaming service for VR end-user devices.

Adaptive streaming systems must deal with issues of the delivery of the tiled 360$^{\circ}$ video from two different perspectives, the service provider and the client. Most recent work focused on the client's perspective~\cite{myIcip2017, Hosseini2016kv, Corbillon2016js,LeFeuvre2016pg, Graf2017, Frounhoundatedwi} without considering the service providers' perspective. More clearly, they neither provide 360$^{\circ}$ video content-specific encoding ladders nor consider the resource costs of the content delivery network (CDN), which is a cloud-based video streaming system that delivers videos to the edge servers so as to effectively connect to the end-users. Given the different characteristic of the tiled 360$^{\circ}$ video content (\textit{e.g.}, ERP and tile encoding), recommended encoding ladders for traditional videos~\cite{appleRec,netflixblogPeerTitle}, that are currently used for adaptive streaming systems, might not achieve an acceptable quality of experience (QoE)~\cite{6982310,janEncoding} for the tiled 360$^{\circ}$ video. Using such encoding ladders might also waste CDN resources and the end-users' bandwidth.
% because the allocated bitrate for a tile \textcolor{red}{might go beyond or above what is necessary in some cases to achieve a noticeable improvement in video quality.}

Our work aims to improve the performance of adaptive 360$^{\circ}$ video streaming systems, providing guidelines for the design of optimal 360$^{\circ}$ VR video streaming systems using tiles. To this end, we focus on the configuration of cost-optimal encoding ladders in adaptive streaming systems by considering both the provider's and client's perspective and develop an encoding ladder estimation method for tiled 360$^{\circ}$ video streaming, which is the main contribution of this work. To the best of our knowledge, such encoding ladder estimation method has not been studied yet. The proposed method deals with minimizing the distortion of the observed tiled 8K$\times$4K ERP video content on the client side while reducing the resource costs on the service provider side, such as storage capacity utilization and computational costs for encoding. In this context, we categorize the given 360$^{\circ}$ videos using their extracted features of encoding complexity, estimate their visual distortion based on the developed distortion model, and calculate the resource costs using the proposed cost models. The cost-optimal encoding ladder configuration problem is then solved using the formulated integer linear programming (ILP) algorithm by considering practical constraints. Our evaluations show that the proposed cost-optimal encoding ladders using a set of service provider's constraint parameters achieve better results compared to the recommended encoding ladders in terms of objective quality for 360$^{\circ}$.

The remainder of this paper is organized as follows. Related work is detailed in Section~\ref{releatedWork}. Then, the proposed system model is presented in Section~\ref{system}. Experiments to demonstrate the performance of our proposed method are presented in Section~\ref{experiment}. Finally, Section~\ref{conclusion} concludes this paper with a summary and future work.

% \import{}{texts/intro}
% % \vspace{-0.8em}
\section{Related Works}
% % \vspace{-0.5em}
\label{releatedWork}
To define the most suitable encoding ladder for traditional video, an unique encoding ladder for each given video content is generated for instance by the engineers at Netflix using the brute-force search algorithm~\cite{netflixblogPeerTitle}. In their research work, each quality-resolution pair was plotted for a given content at each bitrate level. An upper convex hull of its RD curve was then selected to define the encoding ladder. Their approach is very effective concerning QoE for traditional video content. However, it is neither cost-optimal in the sense of resource consumption of a CDN nor content-specific and optimized for tiled 360$^{\circ}$ videos.

% YouTube's work in \cite{chen2016subjective} focused on the bitrate level to switch between 2K and 4K classical videos, and the result showed that the bitrate level should be between 4 and 5 \textit{Mbps}. In the analysis, quality scores were calculated for uploaded videos in YouTube and based upon the calculated scores, and the average switching bitrate between 2K and 4K was defined.

Similarly, academic researchers demonstrated that the previously defined fixed encoding ladders such as Apple's and Netflix's one-size-fits-all schemes~\cite{appleRec,netflixblogPeerTitle}, have critical weaknesses for traditional video content as described in~\cite{Toni_2014_2557652}. Here, the authors defined an optimal encoding ladder for each video category to improve the performance of adaptive streaming for traditional videos. The problem was formulated as an optimization algorithm to find the best bitrate ladder for the given videos by considering the characteristics of a set of end-users in a given database without considering encoding and storage costs. The results have shown, however, that the fixed encoding ladders cannot provide the best objective quality for given traditional videos and clients' bandwidth.

% Along a similar theme, more recent work in \cite{Toni_2014_2557652} explores optimum encoding ladders for the multi-view video content in a scenario of interactive video streaming. \textcolor{red}{the difference}
Most recent work focused on 360$^{\circ}$ video streaming solutions using tiles in order to optimize the quality on the client side~\cite{Frounhoundatedwi, myIcip2017, Hosseini2016kv, Corbillon2016js, LeFeuvre2016pg, Graf2017}. The authors in~\cite{myIcip2017} proposed a new adaptive streaming system based on tiling, integration of the DASH standard and a viewport-aware bitrate level selection method. In~\cite{Hosseini2016kv}, an adaptive bandwidth-efficient 360 VR video streaming system using a divide and conquer approach was presented. The work is based on a dynamic viewport-aware adaptation technique using tiles, derived from a hexaface sphere, and the DASH standard. Similar to the previous work, the authors of~\cite{Corbillon2016js} also propose a viewport-adaptive video delivery system using tiles (cube maps) and different video representations that differ by their bitrate and different scene regions. Additionally, in~\cite{LeFeuvre2016pg}, high-resolution video content is transmitted in tiled fashion using fixed rectangular tiles. The authors in~\cite{Graf2017} presented a bandwidth efficient adaptive 360$^{\circ}$ video streaming system. The work in~\cite{Frounhoundatedwi} described the bandwidth problem of 360$\degree$~video, and suggested to use tile-based streaming. Furthermore, their work described the principles of adaptive streaming of 360$\degree$~video using tiles and evaluated their system with respect to bitrate overhead, bandwidth, and quality requirements. However, none of these works are dealing with cost-optimal encoding ladders on the service provider's side to reduce storage capacity utilization and computational costs.

% Designing an optimal encoding ladder for VR streaming applications that use tiles still remains an unexplored area. To the best of our knowledge, existing 360$^{\circ}$ video streaming solutions lack to use content-specific and cost-effective encoding ladders for 360$^{\circ}$ videos that consider both the client's and the service provider's perspective, respectively.

% we developed a system for VR that supports tiled and viewport-aware adaptive streaming. We used unequal tile sizes and viewport-aware bitrate \textit{distribution} using a novel \textit{distance} criterion. To verify our method, we recorded \textit{real} viewport trajectories from subjects in viewing sessions. With using our recorded data, we calculated the viewport quality scores and compared our proposed method with the reference solution, which is based on the existing professional adaptive streaming systems~\cite{std360industry}.

\section{Proposed System Model}
% \vspace{-0.8em}
\label{system}
\begin{figure*}
\resizebox{\linewidth}{!}{

        \centering
        % \includegraphics[trim={0.2cm 0.2cm 0cm 0.2cm},clip,width=\linewidth]{figures/Add_fig.png}
        %\includestandalone[mode=buildnew,width=1\linewidth]{figures/tikz/overview}
        \includegraphics[width=1\linewidth]{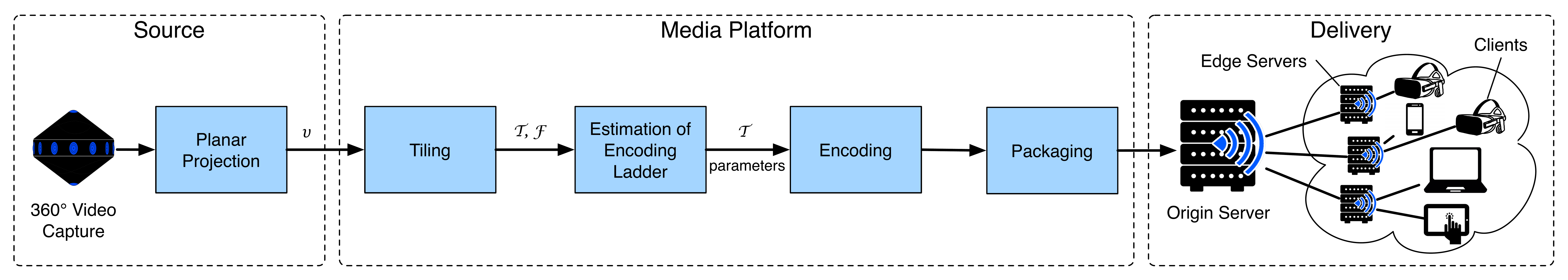}}
        \vspace{-0.5em}
        \caption{Schematic diagram of a cloud-based video streaming pipeline for VR which includes source, media platform, and delivery of the tiled 360$^{\circ}$ video content.}
        \label{mainConcept}
\end{figure*}%
We consider a cloud-based video-on-demand 360$^{\circ}$ video streaming pipeline for VR as depicted in Fig.~\ref{mainConcept}. Each captured 360$^{\circ}$ spherical video is mapped to the ERP representation in 8K$\times$4K resolution for encoding purposes at the source node. The media platform divides each ERP video into $N$ tiles and estimates an unique cost-optimal encoding ladder. Each tile is then encoded at various bitrate levels using multiple encoders with estimated cost-optimal encoding ladder parameters. Then, the generated bitstreams are divided into a set of chunks with equal playback duration, encapsulated by the packaging node and eventually stored on the origin server. Each stored content is then deployed to the CDN, where the bitstreams are efficiently distributed to the VR end-users through the edge servers. 

Each end-user device contains the tiled DASH-VR player \cite{myIcip2017} to communicate with the edge servers and to request individual tiles with appropriate bitrate levels and resolutions from the encoding ladder depending on the bandwidth availability of the network. For adaptive streaming purposes, a set of tiles is encoded at the media platform using different encoding settings. More precisely, let $v$ be an 8K$\times$4K ERP 360$\degree$ video in the set of videos $\mathcal{V}$. Each $v$ is split into $N$ tiles, each tile $j$, $j\in\mathcal{T}$, is then encoded at a different bitrate $b_j$ and resolution $r_j = w_j \times h_j$. Hence, the quadruple ($v$,$j$,$b$,$r$) corresponds to a representation of the video $v\in\mathcal{V}$ for the tile $j\in\mathcal{T}$, encoded at a target bitrate $b\in\mathcal{B}$ and spatial resolution $r\in\mathcal{R}$. Note that $v$, $j$, $b$, and $r$ are integer values and represent the indices of their corresponding sets.

In this context, encoding and accumulating all combinations of the quadruple ($v$,$j$,$b$,$r$) might be very expensive for service providers. Therefore, a cost-effective optimization is required in order to minimize the service provider's resource costs while providing cost-optimal and high quality 360$\degree$ video streaming experience.  
% To this end, we formulate the configuration of the encoding ladder as an optimization problem using the ILP, and propose new encoding ladder estimation method. 

For this aim, the proposed estimation method contains four main parts: classification of the content type, distortion modeling, cost modeling, and problem formulation. First, we extract spatial and temporal features ($f_{spa}$ and $f_{tmp}$) of the $v$-th video to classify its content type as described in subsection~\ref{esEncComp}. Then, we perform an automatic estimation procedure for the encoding ladder using distortion and cost models for the tiled $v$-th video as detailed in subsections~\ref{disModel} and~\ref{costModel}, respectively. Again, in this encoding ladder estimation process we consider both the client side (quality distortion) and service provider side (resource costs). Finally, we formulate the cost-optimal estimation problem for the encoding ladder by applying certain practical constraints, which is eventually solved using the proposed ILP algorithm as described in subsection~\ref{probformul}.

\label{method}
\subsection{Classification of the content type}
\label{esEncComp}

To classify the content type from a given set of content types $\mathcal{O}$, spatial $f_{spa}$ and temporal $f_{tmp}$ complexity features are extracted from the videos. As each video $v$ has different RD performances at various resolutions, we can identify two sources of video distortion: spatial down-sampling and quantization. As a down-sampled version of $v$ suffers from spatial information loss, the level of information loss depends on the \emph{spatial complexity} of each video, which is one of the encoding complexity features. Moreover, the high-resolution version of a given $v$ requires a larger amount of bits to reduce its visual distortion. Compared to its low-resolution version, the high-resolution version has a higher sensitivity for unpredictable motions, which requires further residuals to avoid visual distortions. Since predicted residuals are compressed through quantization which results in quality distortions, \emph{temporal complexity} is the second encoding complexity feature. The content type $o$ of each video is then determined from a given $\mathcal{O}$ by classification using the extracted two complexity features. 

To extract the feature set $\mathcal{F} = \{f_{spa}, f_{tmp}\}$, we use the constant rate factor (CRF) encoding. The CRF encoding, unlike the constant quantization parameter (QP)-based encoding, has the QPs slightly varied across the time based on the scene complexity, action, and motion. For instance, when a scene contains a lot of action and motion, a higher compression can be applied by raising the QP in order to save bitrates. Therefore, the feature set $\mathcal{F}$ can be extracted from the CRF encoded stream to identify the encoding complexity of each $v$. For this purpose, the average size of I- and P- frames can be used as main indicators to determine the complexity features. As also demonstrated in~\cite{chen2016subjective}, the size of I-frames expresses the spatial complexity of each $v$. Thus, we use the normalized version of the I frame sizes to estimate $f_{spa}$ for a given video. As the average size of P frames characterizes the amount of residual bits, we use the ratio of the size of P frames over the size of I-frames as the indicator for $f_{tmp}$. 

\subsection{Distortion modeling}
\label{disModel}

To model the distortion of a given $v$, we model two sources of artifacts, the compression and spatial scaling artifacts, of the tiled 360$\degree$ video using its content type and encoding resolution. Both artifacts, which are the most important distortions that deteriorate QoE, are driven by the encoding target rate and the adaptation of the video resolution to the target resolution. With the aim of reducing search complexity, we generate a continuous distortion model for each content type, as the given parameter space is too large for a the brute-force search algorithm (\textit{e.g.,} Netflix's work in~\cite{netflixblogPeerTitle}). To this end, we derive a distortion function by fitting the two-term power series model using the following fit function:

\begin{equation}
\label{fitfunc}
FT_{ogB} = k_{og} Z_{B}^{\Omega_{og}} + \Phi_{og},
\end{equation}
where $k$, $\Omega$, and $\Phi$ are fitting parameters used in the curve fitting operation for the $o$-th content type, $o\in\mathcal{O}$ and $\mathcal{O}=\{ o_1, o_2, \ldots, o_{|\mathcal{O}|} \}$, of the $g$-the resolution, $g\in\mathcal{G}$ and $\mathcal{G}=\{ g_1, g_2, \ldots, g_{|\mathcal{G}|} \}$, at the tiled ERP video bitrate $B$. Note that $Z$ is the value of the total bitrate of the tiled 360$\degree$ video in terms of \textit{Mbps} (\textit{i.e.,} total bitrate of the ERP video recomposed of the tiles with bitrate $B$). These parameters for the proposed distortion model, shown in Table~\ref{fittingParameters}, were found using the curve fitting operator. Note that index number of $o$ and $g$ are listed in ascending order of their size. The target resolution size is 8K$\times$4K. For the sake of simplicity and also a lack of variety of 8K 360$\degree$ video content types, we only distinguish between three content types and resolutions. Each row and column number of the fitting parameters in the table represents a different content type and resolution, respectively.

\begin{table*}[ht]
 \centering
\footnotesize
\resizebox{\linewidth}{!}{%
      \begin{tabular}{llcccccccccccccccccc}
      \toprule 
    \textbf{Resolution} $\mathcal{G}$ & &\multicolumn{6}{c}{$g_1$} &\multicolumn{6}{c}{$g_2$} &\multicolumn{6}{c}{$g_3$}\\
    \cmidrule(l){3-8}
    \cmidrule(l){9-14}
    \cmidrule(l){15-20}
    \textbf{Model} & &\multicolumn{3}{c}{Distortion} &\multicolumn{3}{c}{Data size} &\multicolumn{3}{c}{Distortion} &\multicolumn{3}{c}{Data size} &\multicolumn{3}{c}{Distortion} &\multicolumn{3}{c}{Data size}\\
    \cmidrule(l){3-5}
    \cmidrule(l){6-8}
    \cmidrule(l){9-11}
    \cmidrule(l){12-14}
    \cmidrule(l){15-17}
    \cmidrule(l){18-20}
    & &$k$ &$\Omega$ &$\Phi$ &$k$ &$\Omega$ &$\Phi$ &$k$ &$\Omega$ &$\Phi$ &$k$ &$\Omega$ &$\Phi$ &$k$ &$\Omega$ &$\Phi$ &$k$ &$\Omega$ &$\Phi$\\
    \multirow{3}{*}{\textbf{Content type} $\mathcal{O}$} 
    &$o_1$ &1809  &-0.6959 & 5.649 &0.7613 &0.9901 &52.54 &4002 & -0.7558 & 2.723 &0.8005 &0.9859 &52.25 & 1829 & -0.5587 & -3.266 &0.8264 &0.9846 &214.9\\
    % & & 1829 & 4002 & 1809 &-0.5587 & -0.7558 & -0.6959 & -3.266  & 2.723 & 5.649\\
    &$o_2$ &220.1 & -0.3583 & 6.447 &0.6467 &1.003 &29.36 &191.9 & -0.2763 & -5.728  &0.6078 &1.009 &71.15 & 480.6 & -0.3643 &-5.728 &0.5654 &1.015 &269\\
    % & & 480.6 &  191.9 &220.1 &-0.3643 & -0.2763 & -0.3583 & -5.728 & -2.552 & 6.447\\
    &$o_3$ &820.4 &-0.4702  &6.2 &0.6631 &1.001 &10.69 &643 &-0.3825 &-2.625 &0.6691 &1 &17.46 & 616.9 &-0.2837 &-23.78 &0.5943 &1.012 &203.8\\
    % & & 616.9 & 643 &820.4 &-0.2837 & -0.3825 & -0.4702 &-23.78 & -2.625 &6.2\\
    \bottomrule
    \end{tabular}
   }
    \caption{Curve fitting parameters for the proposed distortion and data size estimation models.}
    \label{fittingParameters}
    \vspace{-0.8em}
  \end{table*}

To better reflect the distortion of the 360$\degree$ video at the clients' side, we estimate the distortion, caused by the mapping of the spherical content onto the planar surface of the devices (\emph{spherical distortion}), of the tiled 360$\degree$ video as a target value in the curve fitting using the \emph{weighted-to-spherically-uniform mean square error} (WS-MSE)~\cite{Jvet2016tw}. WS-MSE measures the spherical surface using a non-linear weighting in the MSE calculation. Such weights are calculated using the stretching ratio of the area that is projected from the planar surface to the spherical surface. The noise power for the $i$-th representation of the $j$-th tile, $d_{ij}$, can be formulated as follows:
% \vspace{-0.5em}
\begin{equation}
% \vspace{-0.3em}
d_{ij} = \frac{\underset{\substack{x \in W}}{\sum}\underset{
\substack{y \in H}}{\sum} \left( ( t_{j}(x,y) - \tilde{t}_{ij}(x,y) )^2 q_{j}(x,y)\right)}{ \underset{\substack{x \in W}}{\sum}\underset{\substack{y \in H}}{\sum} q_{j}(x,y)} ,
\end{equation}
% \vspace{-0.2em}
where $W \times H$ is the resolution of the reconstructed version of the ERP 360$\degree$ video. Note that $x$ and $y$ denote the pixel coordinates of the ERP video, $t$ and $\tilde{t}$ stand for the original (\textit{i.e.,} uncompressed) and reconstructed versions of the $j$-th tile and $q_j(x,y)$ represents the weighting intensity in ($x$, $y$) of the weight distribution of the ERP for $t_{j}$ which can be calculated according to \cite{Jvet2016tw} with:
% \vspace{-0.5em}
\begin{equation}
% \vspace{-0.3em}
% \label{weightEq}
q_j(x,y) = cos \frac{(y+0.5-H/2)\pi}{H}.
\end{equation}

\subsection{Cost modeling}
\label{costModel}
% \vspace{-0.5em}

In this subsection, we develop cost models for the cloud-based video streaming system in order to minimize the resource costs for encoding workload and storage capacity utilization at the service providers' side. 

\subsubsection{Encoding cost}
The encoding cost is one of the most expensive computing costs which usually occurs on the cloud servers and which heavily depends on the video resolution. To calculate encoding costs, we consider the \textit{broken-line model} where the same cost is defined for similar resolutions. To this end, we extend the cost calculation model used by the Amazon cloud service \cite{amazonPrice} in order to consider broad range of resolution sizes. The encoding cost $c^e$ can be described for the $j$-th tile of the $i$-th representation as follows:

% \vspace{-0.5em}
\begin{equation}
% \vspace{-0.3em}
c^{e}_{ij} = 
\begin{cases}
         \mu_e, &r_{ij}\leq 720p\\
         2\mu_e, &720p< r_{ij}\leq 1080p\\
         4\mu_e, &1080p< r_{ij}\leq 4K\\
         8\mu_e, &4K < r_{ij}\leq 8K\\
\end{cases}
\end{equation}
where $\mu_e$ is a constant term for the encoding cost defined by the service provider and $r_{ij}$ is the resolution of the $j$-th tile in the $i$-th representation.

\subsubsection{Storage cost}

Additionally, large storage capacity is required to store all encoded tiles with different representations for adaptive streaming on the server. The storage cost depends on the data size of the tiled 360$\degree$ video which is located on the server. Considering a linear cost model where the cost is proportional to the data size of each tiled 360$\degree$ video stream, the storage cost $c^s$ for the $j$-th tile of the $i$-th representation can be described as follows:
\begin{equation}
c^s_{ij} = \mu_s bs_{ij} ,
\end{equation}
where $\mu_s$ is a constant term for storage cost defined by the service provider and $bs_{ij}$ is the estimated data size of the $j$-th tile in the $i$-th representation. The data size for each $j$ tile is estimated using the curve fitting technique similar to the one used for Eq.~\eqref{fitfunc}. Parameters for the equation, shown in Table~\ref{fittingParameters} (Data Size), were found using the curve fitting operator.

\subsection{Problem formulation}
\label{probformul}

In order to obtain the cost-optimal encoding ladder $\mathcal{L}^*$ for a given video, a set of representations for $\mathcal{L}^*$ is chosen from the set of the estimated representation $\mathcal{L}$ that minimize both the total spherical quality distortion of tiles and the total resource cost of the cloud-based streaming system. For this purpose, we formulate the problem as an optimization problem using the following practical constraints:
% \vspace{-0.5em}
\begin{enumerate}[label=(\Roman*)]
% \begin{itemize}
    \item \textbf{Bandwidth:} In the proposed system, we consider that the encoding ladder needs to cover a set of given network bandwidth profiles $\mathcal{P} = \{ p_1, p_2, \ldots, p_{|\mathcal{P}|} \}$ with their minimum $B^{min}$ and maximum $B^{max}$ bandwidth ranges.
    % \vspace{-0.5em}
    \item \textbf{Computational and storage costs:} We set limits for the encoding and storage costs which are the maximum allowed computational cost $C^{max}$ and storage cost $S^{max}$ of the streaming system.
% \vspace{-0.5em}
    \item \textbf{Encoding rate:} The bitrate levels of the representations should be spaced between each other by the minimum step size $\tau$.

\end{enumerate}

Our objective is to provide a low-quality distortion encoding ladder for a given tiled $v$ at minimum resource costs by considering the above described constrains. Thus, we formulate the optimization problem as follows:
% \vspace{-0.5em}
\begin{equation}
% \vspace{-0.3em}
\label{ilp1_0}
\mathcal{L}^* : \argmin_{\mathcal{L}} \sum_{i\in \mathcal{L}} \sum_{p\in \mathcal{P}} \left(\gamma c_{i} + (1-\gamma) d_{i}\right) a_{ip}
\end{equation}
with
\begin{equation}
\label{ilp1_1}
c_{i} = \sum_{j \in \mathcal{T}}(c^e_{ij} + c^s_{ij}) \quad  c_i \in \mathcal{P}
\end{equation}
and
\begin{equation}
\label{ilp1_2}
d_{i} = \sum_{j \in \mathcal{T}}d_{ij},
\end{equation}
where $c_{i}$ and $d_{i}$ are the total resource cost and quality distortion for the $i$-th representation, respectively. In order to have a trade-off between  $c_{i}$ and $d_{i}$, we introduce a pre-defined constant $\gamma \in [0, 1]$ to be assigned by the service-provider. To cover a wide range of network bandwidths, we introduce a set of network bandwidth profiles in the problem definition. The decision variable $a_{ip} = \{0, 1\}$ indicates if the $i$-th bitrate level for the $p$-th profile of a set of network bandwidth profiles $\mathcal{P}$ is included \emph{or} excluded in the encoding ladder for a given $v$.

Equation~\eqref{ilp1_0} minimizes both the overall distortion of the tiled 360$\degree$ video and resource costs of the cloud-based streaming system and is subject to the following constraints:\\
\vspace{-0.8em}
\begin{equation}
\vspace{-0.8em}
\begin{split}
\label{ilp2}
B^{min}_p \leq b_i a_{ip} \leq B^{max}_p \quad \forall
i \in \mathcal{L}\ \text{and}\ \forall p \in \mathcal{P},
% ;\ h = 1, 2, &\ldots, N^{h})
\end{split}
\end{equation}
\vspace{-0.6em}
\begin{equation}
\label{ilp3}
% \centering
% \begin{split}
\sum_{i\in \mathcal{L}} a_{ip} = \lfloor \frac{M \Lambda_p}{\sum_{p \in \mathcal{P}} \Lambda_p} \rfloor \quad \quad \forall p \in \mathcal{P} ,
% ;\ h = 1, 2, &\ldots, N^{h})
% \end{split}
\end{equation}
\vspace{-0.5em}
% \textcolor{blue}{adjacent bit rates should be a $\tau$ factor apart}
\begin{equation}
\label{ilp3v2}
% \centering
% \begin{split}
\sum_{p\in \mathcal{P}} a_{ip} \leq 1 \quad \quad \forall i \in \mathcal{L},
% ;\ h = 1, 2, &\ldots, N^{h})
% \end{split}
\end{equation}
\vspace{-0.5em}
% \textcolor{red}{not bigger than storage constraint:}
\begin{equation}
\label{ilp5}
\sum_{i\in \mathcal{L}} \sum_{p\in \mathcal{P}} s_{i} a_{ip} \leq S^{max},
\end{equation}
\vspace{-0.5em}
\begin{equation}
\label{ilp6}
\sum_{i\in \mathcal{L}} \sum_{p\in \mathcal{P}} c_{i} a_{ip} \leq C^{max},
\end{equation}
\vspace{-0.5em}
\begin{equation}
\label{ilp4}
 \frac{b_{i}a_{ip}}{b^*_{n}} \geq \tau ,\quad \forall
i \in \mathcal{L},\  \forall
n \in \mathcal{L^*}\ \text{and}\  \forall p \in P.
\end{equation}
% \begin{equation}
% h_{p} = \frac{P k_p}{\sum_{p \in \mathcal{P}} k_p} \quad (p=1, 2, \ldots P)
% \end{equation}  

Equation~\eqref{ilp2} addresses Constraint (\RNum{1}) for each $p$. Equation~\eqref{ilp3} sets the maximum number of representations in the encoding ladder for the $p$-th profile based on its weighting factor $\Lambda$ and the total number of representations $M$ in the encoding ladder. The weighting factor $\Lambda$ for each network profile is shown in Table~\ref{setupband}. The constraint of Equation~\eqref{ilp3v2} avoids the selection of the same representation for each profile. Additionally, Equations~\eqref{ilp5} and \eqref{ilp6} satisfy Constraint (\RNum{2}) by ensuring that encoded videos for estimated encoding ladders cannot exceed $S^{max}$ and $C^{max}$. Equation~\eqref{ilp4} satisfies Constraint (\RNum{3}) by ensuring that the target bitrate of each selected representation $n$ in the $\mathcal{L}^*$ is spaced by a minimum step size $\tau$.

\section{Experimental Results}
\label{experiment}
% \vspace{-0.5em}

In this section, we investigate the performance of the proposed encoding ladder estimation method by comparison with the one-size-fits-all schemes~\cite{appleRec,axinomRec,netflixblogPeerTitle} for the tiled 360$\degree$~video, and evaluate the proposed method under several service provider's constraints.
% \vspace{-0.5em}
\subsection{Setup}
% \vspace{-0.5em}

We use as the following six 8K$\times$4K resolution 360$\degree$ ERP video test sequences: $\mathcal{V}$ = $\{$\textit{Train}, \textit{Stitched\_left\_Dancing360\_8K}, \textit{Basketball}, \textit{KiteFlite}, \textit{ChairLift}, \textit{SkateboardInLot}$\}$ \cite{AdeelAbbasGoPro,inteldigital,2016rr}. Each $v\in\mathcal{V}$ was split into $N = 10$ tiles which was obtained as an optimal number in our previous research work in~\cite{myIcip2017}. The encoded bitrate for each tile is equally distributed by dividing the \textit{target bitrate} to the $N$ tiles. Their encoding complexity features and assigned content types are shown in Table \ref{setupVideos}, which was estimated using the described method in the Section~\ref{esEncComp}. Three content types in the set, $\mathcal{O}=\{ o_1, o_2, o_3\}$, were used to classify the videos using the estimated complexity features. The \textit{Train}, \textit{Basketball}, and \textit{ChairLift} sequences were used to model the curve fitting function in Equation~\eqref{fitfunc} and we evaluate our method using the \textit{Stitched\_left\_Dancing360\_8K}, \textit{KiteFlite}, and \textit{SkateboardInLot} video sequences. Further, three different resolutions $\mathcal{G} = \{ 3072 \times 1536, 4096 \times 2048, 8192 \times 4096\}$ in the encoding ladders and four different bandwidth profiles $p$ were used as defined in Table~\ref{setupband} with minimum $B^{min}$ and maximum $B^{max}$ bandwidth ranges, and $\Lambda$ for each bandwidth profile.

\begin{table}[htbp] %ht
% \vspace{-0.5em}
\footnotesize
\centering
\begin{tabular}{l c c c c}
\toprule
Profiles: &$p_1$ &$p_2$ &$p_3$ &$p_4$\\
\midrule
\textbf{$B^{min}$} (\textit{Mbps}) &1 &3 &15 &25 \\
\textbf{$B^{max}$} (\textit{Mbps}) &4 &20 &30 &40 \\
$\Lambda$ &0.25 &0.25 &0.25 &0.25\\   
\bottomrule
%\bottomrule
\end{tabular}
\caption{Network bandwidth profiles.}
\label{setupband}
% \vspace{-0.5em}
\end{table}
% \vspace{-0.8em}

We focus on the browser-based video streaming use-case which is one of the core experiments in the ongoing standardization activity~\cite{Jtc1sc29wg2016wm}. Since AVC is the only implemented decoder in current available browsers which can support HMDs, we apply the H.264/AVC standard in our experiments. In this context, we encoded videos using the FFmpeg software (\textit{ver.} N-85291)~\cite{x264} with two-pass and 200 percent constrained variable bitrate encoding configurations. At this stage, it is important to mention that our proposed method is video codec agnostic; it can be easily utilized with different video coding standards.

%We addressed the HMD devices that can render up to 8K resolution ERP 360$\degree$ videos and measured the distortion relative to the original source videos. 

% \subsection{Performance Evaluation}

% \textit{ChairLift}, \textit{SkateboardInLot}, \textit{Basketball}, \textit{Train} JVET-D0026.pdf

% \textit{KiteFlite} JVET-D0039.pdf

\begin{table}[htbp] %ht
% \vspace{-0.5em}
\footnotesize
\centering
\begin{tabular}{l c c c}
%\toprule
% \textbf{Profile} & \textbf{$B_{min}$} (\textit{Mbps}) & \textbf{$B_{max}$} (\textit{Mbps}) & \textbf{Probability}\\ 
\toprule
% k &1 &2 &3 &4\\
Sequence &$f_{spa}$ &$f_{tmp}$ &$\mathcal{O}$\\
\midrule
\textit{Train} &0.977 &0.065 &\multirow{2}{*}{$o_1$}\\ 
\textit{Stitched\_left\_Dancing360\_8K} &0.884 &0.110 &\\ \hline
% \textit{Gaslamp} &402 &A\\
% \textit{JamSession} &566  &A\\  
% \textit{Stitched\_left\_Driving360\_8K} &804 &C\\
% \textit{Trolley} &1008 &C\\
% \textit{Harbor} &942 &C\\
\textit{Basketball} &0.843 &0.090 &\multirow{2}{*}{$o_2$}\\
\textit{KiteFlite} &0.861  &0.090 &\\\hline

\textit{ChairLift} &0.789  &0.212 &\multirow{2}{*}{$o_3$}\\
\textit{SkateboardInLot} &0.827  &0.521 &\\

\bottomrule
%\bottomrule
\end{tabular}
\caption{Encoding complexity features and assigned content types for the used test sequences.}
\label{setupVideos}
% \vspace{-0.5em}
\end{table}
% \vspace{-0.5em}

To evaluate our proposed method, the objective quality metrics WS-MSE and WS-PSNR~\cite{Jvet2016tw} were utilized to calculate the quality performance of the 360$\degree$~video. Further, three different one-size-fits-all encoding ladders (\textit{i.e.,} Apple~\cite{appleRec}, Axinom~\cite{axinomRec}, and Netflix~\cite{netflixblogPeerTitle}), which are recommended for traditional videos, were used as references to investigate the quality performance of our proposed method. Table~\ref{vendorRecom} shows three reference one-size-fits-all encoding ladders for their three ERP resolutions and four total target encoding rates. In the table, resolutions and target encoding rate were calculated by summation of each tile's resolution and target encoding rate, respectively.

\begin{table*}[!ht]
 \centering
  \footnotesize
    \begin{minipage}{1\linewidth}
      \centering
    %   \caption*{8192$\times$4096}
      \begin{tabular}{cccccc}
      \toprule 
 \multicolumn{2}{c}{\textbf{Apple}~\cite{appleRec}} & \multicolumn{2}{c}{\textbf{Axinom}~\cite{axinomRec}}
    & \multicolumn{2}{c}{\textbf{Netflix}~\cite{netflixblogPeerTitle}}\\
    \cmidrule(l){1-2}
    \cmidrule(l){3-4}
    \cmidrule(l){5-6}
    $Z$ (Mbps) &$W \times H$ &$Z$ (Mbps) &$W \times H$ &$Z$ (Mbps) &$W \times H$\\
    45 & 8192 $\times$ 4096 &45 &8192 $\times$ 4096 &43   &8192 $\times$ 4096 \\   
    30 & 8192 $\times$ 4096 &30 &8192 $\times$ 4096 &30   &4096 $\times$ 2048 \\
    20 & 4096 $\times$ 2048 &21 &4096 $\times$ 2048 &23.5 &4096 $\times$ 2048 \\
    11 & 3072 $\times$ 1536 &12 &3072 $\times$ 1536 &17.5 &3072 $\times$ 1536 \\
    \bottomrule
    \end{tabular}
    \end{minipage}
    \caption{Recommended one-size-fits-all encoding ladders for traditional videos by service providers.}
    \label{vendorRecom}
  \end{table*}
% \vspace{-0.5em}
\subsection{Performance evaluation}
% \vspace{-0.5em}

Encoding ladders for our proposed method have been estimated by solving the formulated ILP algorithm in Section~\ref{probformul} using Pyomo (\textit{ver.} 5.0)~\cite{hart2012pyomo}. We set $\mu_e$ and $\mu_s$ to $0.017$ and $0.023$, respectively. These cost values are same as the real cost values in~\cite{amazonPrice}.

% used the default solver GLPK~\cite{makhorin2008glpk} to solve the formulated algorithm.

\begin{figure*}[!ht]
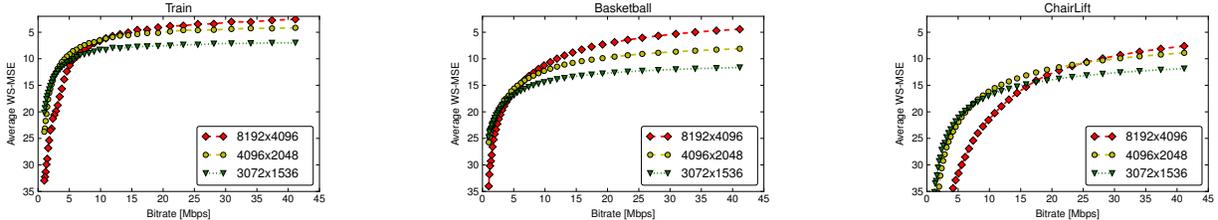

\resizebox{\linewidth}{!}{
        \centering
        \begin{subfigure}[b]{0.38\textwidth}
                \centering
             \scalebox{0.38}{\input{figures/result/Train_cnv.pgf}}
        \end{subfigure}
        \begin{subfigure}[b]{0.38\linewidth}
                \centering
                \scalebox{0.38}{\input{figures/result/Basketball_cnv.pgf}}
        \end{subfigure}
        \begin{subfigure}[b]{0.38\linewidth}
                \centering
                \scalebox{0.38}{\input{figures/result/ChairLift_cnv.pgf}}
        \end{subfigure}
        }
        \caption{Average WS-MSE - bitrate curves for sample 8K$\times$4K ERP 360$\degree$~videos with different content type.}
        \label{convexHull}  
     \vspace{-0.8em}
\end{figure*}

To derive the distortion function in Equation~\eqref{fitfunc}, we calculated the WS-MSE versus bitrate (in \textit{Mbps}) performance graphs in Fig.~\ref{convexHull} for each resolution of the videos \textit{Train}, \textit{Basketball}, and \textit{ChairLift}. The results demonstrate the various performances due to the high diversity in video content characteristics. As can be seen in the figure, each content type has various content dependencies for each encoding resolution and bitrate. For instance, the \textit{Train} sequence (content type $o_1$), which contains the lowest complex encoding features, achieves a low distortion score compared to content types $o_2$ and $o_3$. Because of such diversity, one-size-fits-all schemes, which are used by almost all research works, cannot provide cost-optimal and high-quality streaming performances for the tiled 360$\degree$~videos.

\textbf{Evaluation I:} To evaluate the RD performance gain of our encoding ladder estimation solution, we compare our proposed method with three different recommended one-size-fits-all schemes of the streaming service providers. As these ladders were estimated without considering constraints, we set $\gamma = 0$ (in order to focus on distortion only) and exclude other constraints in equations between \eqref{ilp2} and \eqref{ilp6} for a fair comparison in this test.

\begin{figure*}[htbp]
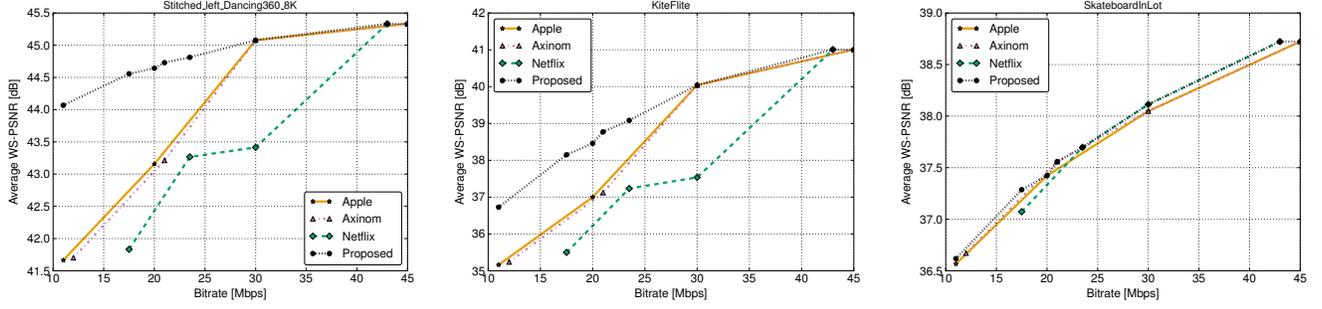

\resizebox{\linewidth}{!}{
        \centering
        \begin{subfigure}[b]{0.33\textwidth}
                \centering
                \scalebox{0.28}{\input{figures/result/Stitched_left_Dancing360_8K_rd.pgf}}
        \end{subfigure}
        \begin{subfigure}[b]{0.33\linewidth}
                \centering
                \scalebox{0.28}{\input{figures/result/KiteFlite_rd.pgf}}
        \end{subfigure}
        \begin{subfigure}[b]{0.33\linewidth}
                \centering
                \scalebox{0.28}{\input{figures/result/SkateboardInLot_rd.pgf}}
        \end{subfigure}}
        \caption{Performance comparison using the RD curves computed with the average WS-PSNR.}
        \label{rdcurves}
\vspace{-0.8em}
\end{figure*}

Figure~\ref{rdcurves} shows the RD curves computed with average WS-PSNR for the \textit{Stitched\_left\_Dancing360\_8K}, \textit{KiteFlite}, and \textit{SkateboardInLot} sequences. The results show that our proposed method considerably increases the objective video quality (\textit{i.e.,} WS-PSNR) compared to the one-size-fits-all schemes at all times. In particular, the proposed method demonstrates high bitrate savings between 10-30 \textit{Mbps} bandwidth ranges for the content types $o_1$ and $o_2$. To this end, we notice that one-size-fits-all schemes provide high scores for the content type $o_3$ compared to their scores for content types $o_1$ and $o_2$.

\textbf{Evaluation II:} We further analyze the performance gain of our method using the Bj{\o}ntegaard metric~\cite{Bjontegaard} in Table~\ref{bdrates}. This metric describes the distance between two RD curves. In this manner, the bitrate difference, \textit{i.e.} BD-rate, was calculated in percentage averaged over the entire range. A negative BD-rate indicates a decrease of bitrate at the same quality. From the table, we can notice that the proposed method provides considerable bitrate savings compared to the recommended encoding ladders at the same bitrates.

\begin{table}[htbp] %ht
\centering
\resizebox{\columnwidth}{!}{%
\begin{tabular}{l c c c}
%\toprule
% \textbf{Profile} & \textbf{$B_{min}$} (\textit{Mbps}) & \textbf{$B_{max}$} (\textit{Mbps}) & \textbf{Probability}\\ 
\toprule
% k &1 &2 &3 &4\\
\multirow{2}{*}{\textbf{Sequence} $v$} &\multicolumn{3}{c}{\textbf{Streaming vendor}}\\
\cmidrule(l){2-4}
&Apple &Axinom &Netflix\\
\midrule
\textit{Stitched\_left\_Dancing360\_8K} &-5.557 &-5.885 &-69.253\\ \hline

\textit{KiteFlite} &-13.876 &-14.436 &-69.178\\\hline

\textit{SkateboardInLot} &-1.673 &-1.701 &-1.155\\

\bottomrule
%\bottomrule
\end{tabular}
}
% \vspace{-1em}
\caption{BD-rate saving (\%) of the proposed method.}
% \vspace{-1em}
\label{bdrates}
\end{table}

\textbf{Evaluation III:} Finally, in the last set of evaluations, we consider a scenario where the constraints of $S^{max}$ and $C^{max}$ are \textit{8000}, $\tau$ = 1.2, and $M$ = 12. In this setup, we use the normalized difference of the total cost $\Delta CS$ and the distortion $\Delta DS$ (in terms of WS-MSE) in percentages for evaluation purpose. Table~\ref{propRes} shows the results of the proposed encoding ladder estimation using resolution-bitrate pairs for $\gamma$ = 0, $\gamma$ = 0.1, and $\gamma$ = 0.5.

\begin{table*}[!ht]
 \centering
\begin{adjustbox}{width=1\textwidth}

      \begin{tabular}{l|l|cccccccccccc}
      \toprule 
\multicolumn{1}{c}{\multirow{2}{*}{Sequence $v$}} &\multicolumn{1}{c}{\multirow{2}{*}{$\gamma$}} &\multicolumn{12}{c}{Representation $i$}\\
& &1 &2 &3 &4 &5 &6 &7 &8 &9 &10 &11 &12\\
\midrule
\multirow{3}{*}{\textit{Stitched\_left\_Dancing360\_8K}} &$0.0$ &($g_1$,1.47) &($g_1$,1.78) &($g_1$,2.15) &($g_1$,3.8) &($g_1$,4.6) &($g_1$,5.6) &($g_2$,10.84) &($g_2$,13.11) &($g_2$,15.87) &($g_2$,28.11) &($g_3$,34.01) &($g_3$,41.15)\\
&$0.1$ &($g_2$,1.34) &($g_2$,1.61) &($g_2$,1.95) &($g_2$,2.60) &($g_3$,3.14) &($g_3$,3.80) &($g_3$,6.12) &($g_3$,7.40) &($g_3$,8.96) &($g_3$,17.45) &($g_3$,21.12) &($g_3$,25.55)\\ %\cmidrule(r){2-5}
&$0.5$ &($g_2$,1.00) &($g_2$,1.21) &($g_2$,1.47) &($g_2$,2.36) &($g_3$,2.86) &($g_3$,3.46) &($g_3$,6.12) &($g_3$,7.40) &($g_3$,8.96) &($g_3$,17.45) &($g_3$,21.12) &($g_3$,25.55)\\ %\cmidrule(r){2-5}
\hline

\multirow{3}{*}{\textit{KiteFlite}}
&$0.0$ &($g_1$,1.47) &($g_1$,1.78) &($g_2$,2.15) &($g_2$,3.80) &($g_2$,4.60) &($g_3$,5.56) &($g_3$,10.84) &($g_3$,13.11) &($g_3$,15.87) &($g_3$,28.11) &($g_3$,34.01) &($g_3$,41.15)\\
&$0.1$ &($g_1$,1.47) &($g_1$,1.78) &($g_2$,2.15) &($g_2$,3.80) &($g_2$,4.60) &($g_3$,5.56) &($g_3$,6.73) &($g_3$,8.14) &($g_3$,9.85) &($g_3$,17.45) &($g_3$,21.12) &($g_3$,25.55)\\ %\cmidrule(r){2-5}
&$0.5$ &($g_1$,1.00) &($g_1$,1.21) &($g_1$,1.47) &($g_2$,2.36) &($g_2$,2.86) &($g_2$,3.46) &($g_3$,6.12) &($g_3$,7.40) &($g_3$,8.96) &($g_3$,17.45) &($g_3$,21.12) &($g_3$,25.55)\\ %\cmidrule(r){2-5}
\hline

\multirow{3}{*}{\textit{SkateboardInLot}}
&$0.0$ &($g_1$,1.47) &($g_1$,1.78) &($g_1$,2.15) &($g_1$,3.80) &($g_1$,4.60) &($g_1$,5.56) &($g_2$,10.84) &($g_2$,13.11) &($g_2$,15.87) &($g_2$,28.11) &($g_3$,34.01) &($g_3$,41.15)\\
&$0.1$ &($g_1$,1.47) &($g_1$,1.78) &($g_1$,2.15) &($g_1$,2.86) &($g_1$,3.46) &($g_1$,4.18) &($g_1$,6.12) &($g_1$,7.40) &($g_1$,8.96) &($g_1$,17.45) &($g_2$,21.12) &($g_2$,25.55)\\ %\cmidrule(r){2-5}
&$0.5$ &($g_1$,1.21) &($g_1$,1.47) &($g_1$,1.78) &($g_1$,2.36) &($g_1$,2.86) &($g_1$,3.46) &($g_1$,6.12) &($g_1$,7.40) &($g_1$,8.96) &($g_2$,17.45) &($g_2$,21.12) &($g_2$,25.55)\\ %\cmidrule(r){2-5}
\hline

    \bottomrule
    \end{tabular}
\end{adjustbox}
% \vspace{-1em}
    \caption{Results of the proposed encoding ladder estimation for $\gamma$ = 0, $\gamma$ = 0.1, and $\gamma$ = 0.5.}
    \vspace{-1em}
    \label{propRes}
  \end{table*}
From the results, we observe that the lowest complex content, \textit{i.e.,} content type $o_1$, increases its encoding resolution and decreases its target encoding rate at the range between $i = 2$ and $i = 10$ to reduce the total cost by considering cost and distortion tradeoffs using $\gamma$ = 0.1 and $\gamma$ = 0.5. On the other hand, we observe that the most complex content, \textit{i.e.} content type $o_3$, decreases both its encoding resolution and target encoding rate in order to reduce the total cost by considering cost and distortion tradeoffs using the $\gamma$ = 0.1 and $\gamma$ = 0.5. Table~\ref{costSaving} reports the total cost saving and distortion gain with respect to different $\gamma$. Finally, we would like to mention that, the GNU linear programming kit (GLPK) for Pyomo was able to solve the formulated ILP algorithm in Section~\ref{probformul} using the calculated data in less than one minute on Intel(R) Core(TM) i7-6700 CPU @ 3.40GHz with 32 GB of RAM.

\begin{table}[htbp] %ht
\centering
  \footnotesize
\resizebox{\columnwidth}{!}{%
\begin{tabular}{l c c c c}
%\toprule
% \textbf{Profile} & \textbf{$B_{min}$} (\textit{Mbps}) & \textbf{$B_{max}$} (\textit{Mbps}) & \textbf{Probability}\\ 
\toprule
% k &1 &2 &3 &4\\
\multirow{2}{*}{\textbf{Sequence} $v$} &\multicolumn{2}{c}{$\Delta$cost (\%)} &\multicolumn{2}{c}{$\Delta$distortion (\%)}\\
\cmidrule(l){2-3} 
\cmidrule(l){4-5}
&$\gamma = 0.1$ &$\gamma =0.5$ &$\gamma = 0.1$ &$\gamma =0.5$\\
\midrule
\textit{Stitched\_left\_Dancing360\_8K} &37.463 &39.683 &-13.628 &-42.914\\ \hline

\textit{KiteFlite} &33.165 &39.206 &-9.564 &-25.326\\\hline

\textit{SkateboardInLot} &37.214 &38.884 &-8.977 &-15.26\\

\bottomrule
%\bottomrule
\end{tabular}
}
% \vspace{-1em}
\caption{Total cost saving and distortion gain with respect to $\gamma$=0.0.}
\vspace{-1em}
\label{costSaving}
\end{table}

% To this end, in our future work, we will evaluate the proposed method using different constraint numbers (\textit{e.g.,} $\gamma$, $M$, $S^{max}$, $C^{max}$) to investigate the effect of encoding and storage costs and visual distortion.

% \vspace{-0.5em}

\section{Conclusions}
% \vspace{-0.8em}
\label{conclusion}

This paper introduced a novel encoding ladder estimation method for tiled 360$^{\circ}$ video streaming systems, considering both the provider's and client's perspectives. To this end, the objective of our proposed method was to provide cost-optimal and enhanced video streaming experiences for VR end-users. The developed system included classification of the content type, distortion modeling, cost modeling, and problem formulation. The performance of our proposed method was verified in experimental evaluations. The results showed that our method achieved significant bitrate savings (especially for the content types $o_1$ and $o_2$) compared to the one-size-fits-all encoding ladders which are recommended by streaming service providers. Furthermore, the developed method can automatically find cost-optimal encoding ladders using several practical constraints, and provides efficient streaming service for tiled 360$^{\circ}$ video. As future work, we plan to extend our optimization framework by considering the number of tiles for a given content type and investigating the effect of total costs by evaluating the effects of the various constraint parameters using a larger set of video sequences.

% investigating tile size optimization and tile discarding, as they are expected to further increase the visual quality. 

% As future work, we plan to extend our optimization by considering the number of tiles for a given content type and user navigation path.
% \bibliographystyle{plain}

\bibliographystyle{IEEEbib}
% \bibliography{Jan03}
% \bibliography{Jan18}

% To the best of our knowledge, this paper is the first study
% about optimal encoding parameters for representation sets
% in free-viewpoint adaptive streaming. We have defined an
% optimization problem for the selection of the representation
% set that maximizes the average satisfaction of interactive users
% while minimizing their view-switching delay. We define a
% novel variable, namely the multi-view navigation segment,
% and formulate an optimization problem that can be solved
% as a tractable ILP problem. We characterize the satisfaction
% of interactive users as the quality experienced by the user
% during the navigation. This function is able to take into
% account both coding and view synthesis artifacts. We finally
% measure the performance of representation sets based on
% content provider recommendations and show the suboptimality
% of baseline algorithms that do not adapt the coding parameters
% to the video and users characteristics. We therefore highlight
% the gap between existing recommendations and solutions that
% maximize the average user satisfaction. In particular, we show
% that an unequal allocation of the storage capacity among
% different video types as well as camera views is essential to
% strike for the right balance between storage cost and users
% satisfaction in interactive multi-view video systems.

{\footnotesize
\bibliography{ref}}

\end{document}